\documentclass[aps, prb, showpacs, twocolumn, amsfonts, amsmath, amssymb, superscriptaddress, floatfix]{revtex4-1}

\usepackage[T1]{fontenc}
\usepackage[utf8x]{inputenc} 
\usepackage{amsmath}
\usepackage{amssymb}
\usepackage{color}
\usepackage{graphicx}
\usepackage{dsfont}

\def\be{\begin{equation}}
\def\ee{\end{equation}}
\def\bea{\begin{eqnarray}}
\def\eea{\end{eqnarray}}

\begin{document}

\title{Absence of two-body delocalization transitions in the two-dimensional Anderson-Hubbard model}
\author{Filippo Stellin} 
\email{filippo.stellin@univ-paris-diderot.fr}
\affiliation{Universit\' e de Paris, Laboratoire Mat\' eriaux et Ph\' enom\`enes Quantiques, CNRS, F-75013, Paris, France}
\author{Giuliano Orso}
\email{giuliano.orso@univ-paris-diderot.fr}
\affiliation{Universit\' e de Paris, Laboratoire Mat\' eriaux et Ph\' enom\`enes Quantiques, CNRS, F-75013, Paris, France}

\date{\today}

\begin{abstract}

We investigate Anderson localization of two particles moving in a two-dimensional (2D) disordered lattice and coupled by contact interactions. Based on transmission-amplitude calculations for relatively large strip-shaped grids, we find that all pair states are localized in lattices of infinite size. In particular, we show that previous claims of an interaction-induced mobility edge are
 biased by severe finite-size effects. 
The localization length of a pair with zero total energy exhibits a nonmonotonic behavior as a function of the interaction strength, 
characterized by an exponential enhancement in the weakly interacting regime. 
Our findings also suggest that the many-body mobility edge of the 2D Anderson-Hubbard model disappears in the zero-density limit, irrespective of the (bosonic or fermionic) quantum statistics of the particles. 
\end{abstract}

\maketitle

\maketitle

\section{Introduction}
It is well known that in certain disordered media wave propagation can be completely halted due to the back-scattering of the randomly distributed impurities.
 This phenomenon, known as Anderson localization~\cite{Anderson:LocAnderson:PR58}, has been reported for different  kinds of waves, such as light waves in diffusive media~\cite{Wiersma:LightLoc:N97,Maret:AndersonTransLight:PRL06} or in disordered photonic crystals~\cite{Segev:LocAnderson2DLight:N07,Lahini:AndersonLocNonlinPhotonicLattices:PRL08}, ultrasound~\cite{vanTiggelen:AndersonSound:NP08}, microwaves~\cite{Chabanov:StatisticalSignaturesPhotonLoc:N00} and atomic matter waves~\cite{Billy:AndersonBEC1D:N08,Roati:AubryAndreBEC1D:N08}.
Its occurrence is ruled by the spatial dimension of the system and by the symmetries of the model, which determine its universality class~\cite{Altland:PRB1997}. 
When both spin-rotational and time-reversal symmetries are preserved, notably in the absence of magnetic fields and spin-orbit couplings, all wave-functions are exponentially localized in one and two dimensions. In three and higher dimensions the system possesses both localized and extended states, separated in energy by a critical point, dubbed the mobility edge, where the system
undergoes a metal-insulator transition~\cite{Evers:AndersonTransitions:RMP08}. 
Anderson transitions have recently been detected using noninteracting atomic quantum gases~\cite{Kondov:ThreeDimensionalAnderson:S11,Jendrzejewski:AndersonLoc3D:NP12,Semeghini:2014} exposed to three-dimensional (3D) speckle potentials. Theoretical predictions for the mobility edge of atoms have also been reported~\cite{Yedjour:2010,Piraud:PRA2014,Delande:MobEdgeSpeckle:PRL2014,Pilati:LevelStats:2015,Pasek:3DAndersonSpeckle:PRA2015,Pilati:3DAndersonSpeckle:2015,Pasek:PRL2017,Orso:SpinOrbit:PRL2017} and compared with the experimental data.

Interactions can nevertheless significantly perturb the single-particle picture of Anderson localization. Puzzling metal-insulator transitions~\cite{Kravchenko:PRB1994}, discovered in high-mobility 2D electron systems in silicon, were later interpreted theoretically 
in terms of a two-parameter scaling theory of localization, which combines  disorder and strong electron-electron interactions~\cite{Punnoose:Science2005,Knyazev:PRL2008}.  
In more recent years a growing interest has emerged 
 around the concept of many-body localization~\cite{GornyiPRL2005,Altshuler:MetalInsulator:ANP06} (MBL), namely the generalization of Anderson localization to disordered interacting quantum systems at finite particle density (for recent reviews see Refs.~\cite{Nandkishore2015,ALET2018498,Abanin:RMP2019}). 
 In analogy with the single-particle problem, MBL phases are separated from (ergodic) thermal phases by critical points situated at finite energy density, known as many-body mobility edges.  
 While MBL has been largely explored in one dimensional  systems with short range interactions,
both experimentally~\cite{Schreiber:Science2015,Rispoli:Nature2019} and
theoretically~\cite{PhysRevB.75.155111,PhysRevB.91.081103,Michal:PRL2014,Andraschko:PRL2014,Mondaini:PRA2015,Reichl:PRA2016,Prelovsek:PRB2016,Zakrzewski:PRB2018,Hopjan:PRA2020,krause2019nucleation,yao2020manybody}, its very existence 
in systems with higher dimensions remains  unclear. 
In particular it has been suggested~\cite{DeRoeck:PRB2016,DeRoeck:PRB2017} that the MBL is inherently unstable against thermalization in large enough samples. This prediction contrasts with subsequent experimental~\cite{Choi1547} and numerical~\cite{WahlNatPhys2019,geiler2019manybody,De_Tomasi_2019,Thomson:PRB2018} studies of 2D systems of moderate sizes, showing evidence of a many-body mobility edge. 
It must be emphasized that thorough numerical investigations, including a finite-size scaling analysis, are computationally challenging beyond one  dimension~\cite{theveniaut2019manybody}.

In the light of the above difficulties, it is interesting to focus on the localization properties of few interacting particles in large (ideally infinite) disordered lattices. 
Although these  systems may represent overly simplified examples of MBL states, 
they can show similar effects, including interaction-induced delocalization transitions with genuine mobility edges\cite{Stellin:PRB2019,stellin2020twobody}. 
In a seminal paper~\cite{Shepelyansky:AndLocTIP1D:PRL94}, Shepelyansky showed that two particles moving in a one-dimensional lattice and coupled by contact interactions can travel over a distance much larger than the single-particle localization length, before being  localized by the disorder. This intriguing effect was confirmed by several numerical studies~\cite{Weinmann:PRL1995,vonOppen:AndLocTIPDeloc:PRL96,Frahm1999,Roemer:PhysicaE2001,Krimer:JETP2011,Dias:PhysicaA2014,Lee:PRA2014,Krimer:InterConnDisord2PStates:PRB15,Frahm:EigStructAL1DTIP16,Thongjaomayum:PRB2019,thongjaomayum2020multifractality}, trying to identify the explicit dependence of the pair localization length on the interaction strength. Quantum walk dynamics of two interacting particles moving in a disordered one-dimensional lattice has also been explored, revealing subtle correlation effects~\cite{Lahini:PRL2010,Chattaraj:PRA2016,Dariusz:PRA2017,Toikka:PRB2020,Malishava:PRB2020}. 
Interacting few-body systems with more than two particles have also been studied numerically in one dimension, confirming the stability of the localized phase. In particular Ref.~\cite{Mujal:PRA2019} investigated  a model of  up to three  bosonic atoms  with mutual contact interactions and subject to a spatially correlated disorder generated by laser speckles, while Ref.~\cite{Schmidtke:PRB2017} addressed
the localization in the few-particle regime of the XXZ spin-chain with a random magnetic field.
 
The localization of two interacting particles has been much less explored in dimensions higher then one. Based on analytical arguments, it was suggested~\cite{Imry:CohPropTIP:EPL95, Borgonovi:NonLinearity1995} that all two-particle states  are localized by the disorder in two dimensions, whereas in three dimensions a delocalization transition for the pair could occur even if all single-particle states are localized.
Nevertheless subsequent numerical investigations~\cite{Ortugno:AndLocTIPDeloc:EPL99,Cuevas:PRL1999,Roemer1999}  in two dimensions reported  evidence of an Anderson transition for the pair, providing explicit results for the corresponding position of the mobility edge and the value of the critical exponent. 

Using large-scale numerics, we recently investigated~\cite{Stellin:PRB2019,stellin2020twobody} 
Anderson transitions for a system of two interacting particles (either bosons or fermions with opposite spins), obeying the 3D Anderson-Hubbard model. We showed that the phase diagram in the energy-interaction-disorder space contains multiple metallic and insulating regions, separated by two-body mobility edges. In particular we observed metallic pair states for relatively strong disorder, where all single-particle states are localized, which can be thought of as a proxy for interaction-induced many-body delocalization.   Importantly,  our numerical data for the metal-insulator transition were found to be consistent with the (orthogonal) universality class of the noninteracting model. This feature is not unique to our model, since single-particle excitations in a disordered many-body electronic system also undergo a metal-insulator transition  belonging to the noninteracting universality class~\cite{Burmistrov:PRB2014}. 

In this work we revisit the Shepelyansky problem in two dimensions and shed light on the controversy. We find that no mobility edge exists for a single pair in an infinite lattice, although interactions can dramatically enhance the pair localization length. In particular we show that previous claims~\cite{Ortugno:AndLocTIPDeloc:EPL99,Cuevas:PRL1999,Roemer1999}  of 2D interaction-driven Anderson transitions 
were plagued by strong finite-size effects. 

The paper is organized as follows. In Sec.~\ref{sec:theory} we revisit the theoretical approach based on the exact mapping
of the two-body Schrodinger equation onto an effective single-particle problem for the center-of-mass motion. 
The effective model allows to recover the entire energy spectrum of orbitally symmetric pair states and is therefore equivalent to the exact diagonalization of the full Hamiltonian in the same subspace; an explicit proof for a  toy 
Hamiltonian is given in Sec.~\ref{sec:equivalence}.
In Sec.~\ref{sec:absence} we present the
finite-size scaling analysis  used to discard  the existence of the  2D Anderson transition for the pair, while in  Sec.~\ref{sec:loclength} 
we discuss the dependence of the two-body localization length on the interaction strength. The generality of the obtained results
is discussed in Sec.~\ref{general} while in Sec.~\ref{sec:conclusions} we provide
a summary and an outlook.
 
 
\section{Effective single-particle model for the pair}
 \label{sec:theory}
The Hamiltonian of the two-body system can be written as $\hat H=\hat H_0 + \hat U$, whose noninteracting part $\hat H_0$ can be decomposed as $\hat H^\textrm{sp} \otimes  \hat{\mathds{1}} +\hat{\mathds{1}}  \otimes \hat H^\textrm{sp}$. Here $\hat{\mathds{1}}$ refers to the one-particle identity operator, while $\hat H^\textrm{sp}$ denotes the single-particle Anderson Hamiltonian:
\begin{equation}
\label{Anderson3D}
\hat H^\textrm{sp}= -J \sum_{\langle \mathbf n, \mathbf m\rangle} |\mathbf m  \rangle  \langle \mathbf n| + \sum_{\mathbf n}V_\mathbf n |\mathbf n\rangle \langle \mathbf n|,
\end{equation}
where $J$ is the tunneling amplitude between nearest neighbor sites $\mathbf{m}$ and $\mathbf{n}$, whereas $V_{\mathbf{n}}$ represents the value of the random potential at site $\mathbf{n}$. 
In the following we consider a random potential which is spatially uncorrelated $\langle V_\mathbf n V_{\mathbf n^\prime} \rangle= \langle V_\mathbf n^2\rangle \delta_{\mathbf n \mathbf n^\prime}$ and obeys a uniform on-site distribution, as in Anderson's original work~\cite{Anderson:LocAnderson:PR58}:
\be\label{randombox}
P(V)=\frac{1}{W}\Theta(W/2-|V|),
\ee
where $\Theta(x)$ is the Heaviside (unit-step) function and $W$ is the disorder strength. The two particles are coupled together by contact (Hubbard) interactions described by
\be\label{intro1}
\hat U=U\sum_{\mathbf m}|{\mathbf m},{\mathbf m}\rangle \langle {\mathbf m},{\mathbf m}|,
\ee
where $U$ represents the corresponding strength. We start by writing the two-particle Schr{\"o}dinger equation as $(E -\hat H_0)|\psi\rangle=\hat U|\psi\rangle$, where $E$ is the total energy of the pair. 
If  $U|\psi\rangle =0$, then $E$ must belong to the energy spectrum of the
noninteracting Hamiltonian $\hat H_0$. This occurs for instance if the two-particles correspond to fermions in the spin-triplet state, as in this 
case the orbital part of the wave-function is antisymmetric and therefore
$\langle {\mathbf m},{\mathbf m}|\psi\rangle=0$.

Interactions are instead relevant for orbitally symmetric wave-functions, describing either bosons or fermions with opposite spins in the singlet state.
In this case from Eq.~(\ref{intro1}) we find that the wave-function obeys the following self-consistent equation
  \begin{equation}
\label{formalism2}
|\psi\rangle=\sum_{\mathbf m} U \hat G(E) |{\mathbf m},{\mathbf m}\rangle \langle {\mathbf m},{\mathbf m}|\psi\rangle,  
\end{equation}
where $\hat G(E)=(E \hat I -\hat H_0)^{-1}$ is the non-interacting two-particle Green's function. Eq.~(\ref{formalism2}) shows that
for contact interactions the wave-function of the pair can be completely determined once its diagonal amplitudes
$f_{\mathbf m}=\langle {\mathbf m},{\mathbf m}|\psi\rangle$ are known.
 By projecting Eq.(\ref{formalism2}) over the state 
$|{\mathbf n},{\mathbf n}\rangle$, we see that these terms obey a closed equation~\cite{Stellin:PRB2019,Dufour:PRL2012,Orso:PRL2005}: 
 \begin{equation}
 \label{integral}
\sum_{\mathbf m} K_{\mathbf n  \mathbf m} f_{\mathbf m} = \frac{1}{U}f_{\mathbf n},
 \end{equation} 
where  $K_{\mathbf n  \mathbf m}  =\langle {\mathbf n},{\mathbf n }|\hat G(E) |{\mathbf m},{\mathbf m}\rangle$. Eq.(\ref{integral}) is then interpreted as an effective single-particle problem with Hamiltonian matrix $K$ and pseudoenergy $\lambda=1/U$, corresponding to the inverse of the interaction strength. 
In the following we will address the localization properties of this effective
model for the pair.
To this respect, we notice that the matrix elements of $K$ are unknown and must be calculated explicitly in terms of the  eigenbasis of the single-particle model, $\hat H^\textrm{sp} | \phi_r\rangle=\varepsilon_r  | \phi_r\rangle$, as
\begin{equation}\label{KE0}
K_{\mathbf n  \mathbf m} = \sum_{r,s=1}^N \frac{\phi_{\mathbf n r}  \phi_{\mathbf m r}^*  \phi_{\mathbf n s}  \phi_{\mathbf m s}^*}{E-\varepsilon_r-\varepsilon_s}, 
\end{equation}
where $N$ is the total number of lattice sites in the grid and $\phi_{\mathbf n r} =\langle \mathbf n | \phi_r\rangle$ are the amplitudes of the one-particle wave-functions. 
%

\section{Equivalence with exact diagonalization of the full model}
\label{sec:equivalence}
\begin{figure}
	\centering
	\includegraphics[width=\columnwidth]{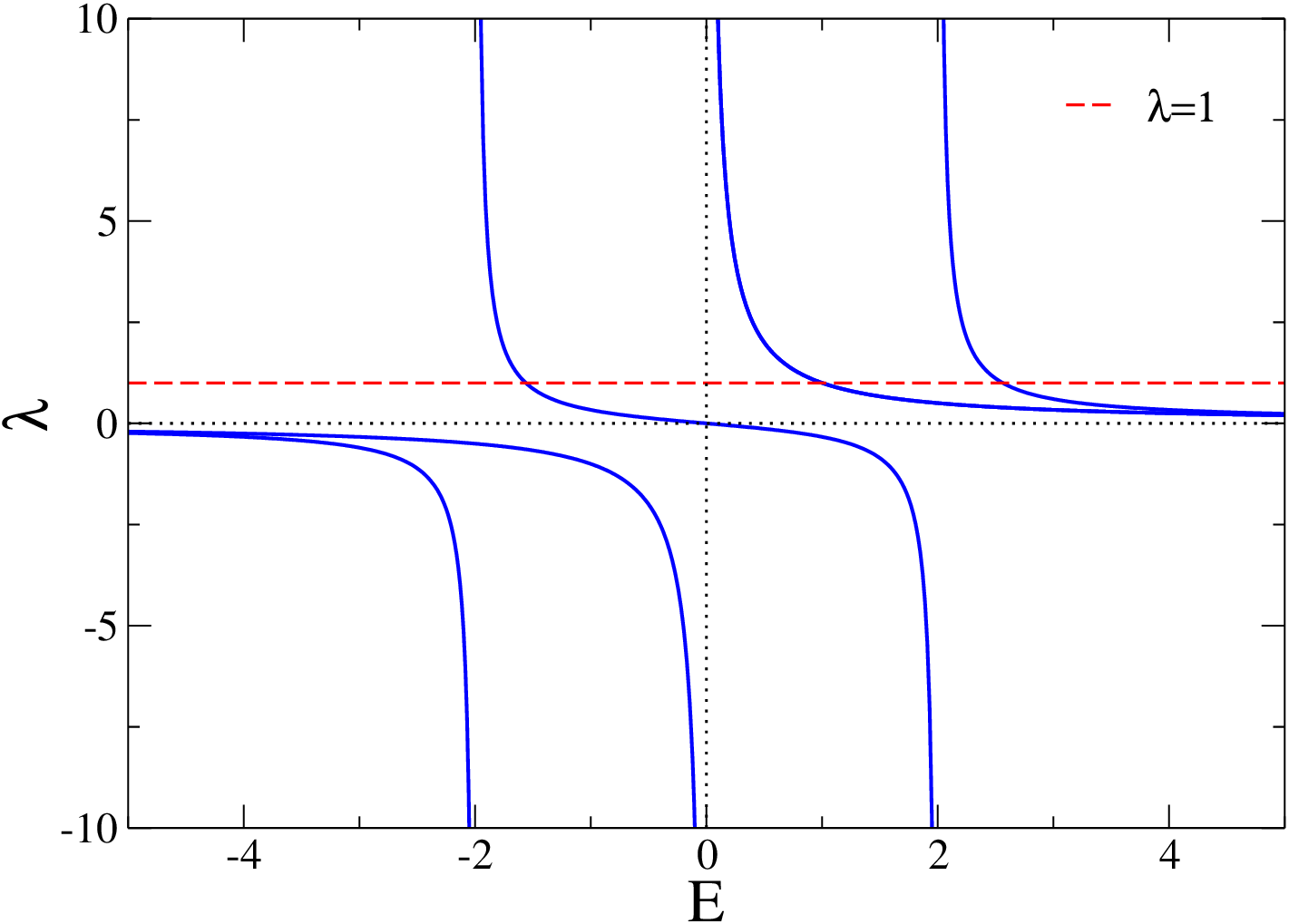}
	\caption{Eigenvalues of the matrix $K$ of the effective model of the pair, Eq.~(\ref{integral}) for a toy model of  $N=2$ coupled sites with no disorder, plotted as a function of the energy $E$ of the pair (blues data curves).
	For a given interaction strength $U$, the entire spectrum of $N(N+1)/2$ energy levels of orbitally symmetric states of the pair can be obtained by intersecting the data curves with the horizontal line, $\lambda=1/U$, here shown for $U=1$ (dashed red line). The corresponding three energy levels are $E=-1.56155$, $E=1$ and $E=2.56155$. }
	\label{fig:example}
\end{figure}

The effective single-particle model of the pair, Eq.~(\ref{integral}), allows to 
reconstruct the entire energy spectrum of orbitally symmetric states for a given interaction strength $U$. 
At first sight this is not obvious  because the matrix $K$ is $N\times N$, and therefore possesses $N$ eigenvalues, while the dimension of the Hilbert space of orbitally  symmetric states is $N(N+1)/2$, which is much larger.  
The key point is that one needs to compute the matrix $K$ and the associated eigenvalues $\lambda_{r}=\lambda_{r}(E)$, with $r=1,2 ...N$, for different values of the energy $E$. The energy levels for fixed $U$ 
are then obtained by solving the equations $\lambda_{r}(E)=1/U$ via 
standard root-finding algorithms. 
Let us  illustrate the above point for a toy model with $N=2$ lattice sites in the absence of disorder. 
In this case the Hilbert space of symmetric states is spanned by the  three vectors $|1,1\rangle$,
$|2,2\rangle $ and $(|1,2\rangle +|2,1\rangle)/\sqrt{2}$. 
The corresponding energy levels of the pair can be found from the exact diagonalization of the $3\times 3$ matrix of the projected Hamiltonian:
\be
H_{ed}=   
 \begin{pmatrix}
	U & -\sqrt{2} & 0  \\
	-\sqrt{2} & 0 &  -\sqrt{2} \\
	0 & -\sqrt{2} & U 
	\end{pmatrix}.
\ee
An explicit calculation yields $E=U$ and $E=(U\pm \sqrt{U^2+16})/2$.
Let us now show that we recover exactly the same results using our effective model. 
The single-particle Hamiltonian is represented by the matrix 
\be\label{example}
H^{sp}=\begin{pmatrix}
	0 & -1\\
	-1 & 0
\end{pmatrix},
\ee
whose eigenvalues are given by $\varepsilon_1=-1$ and $\varepsilon_2=1$. The associated wavevectors are $| \phi_1\rangle =(|1\rangle +|2\rangle)/2$ and
$| \phi_2 \rangle =(|1\rangle -|2\rangle)/2$.
From Eq.(\ref{KE0}) we immediately find 
 \be\label{example2}
 K=\begin{pmatrix}
 	A & B\\
 	B & A 
    \end{pmatrix},
 \ee
where $A=(E/(E^2-4)+1/E)/2$ and $B=(E/(E^2-4)-1/E)/2$. The corresponding eigenvalues of $K$ are given by $\lambda_1(E)=A-B=1/E$ and $\lambda_2(E)=A+B=E/(E^2-4)$. The condition $\lambda_1=1/U$ yields $E=U$, while
$\lambda_2=1/U$ admits two solutions, $E=(U\pm \sqrt{U^2+16})/2$, allowing to recover the exact-diagonalization energy spectrum.
In Fig.\ref{fig:example} we plot the energy dependence of the two eigenvalues of $K$ for our toy model. Intersecting the curves 
with the horizontal line $\lambda=1/U$ (dashed red line) yields visually the three sought energy levels for the orbitally symmetric states.
%

We stress that extracting the full energy spectrum of the pair based on the effective model, for a fixed value of the interaction strength $U$, is computationally demanding as $N$ becomes large.
The effective model is instead very efficient, as compared to the exact diagonalization, when we look at the properties of the pair as a function of the interaction strength $U$, for a fixed value of the total energy $E$. This is exactly the situation that we will be interested in below.

\section{Absence of 2D delocalization transitions for the pair}
\label{sec:absence}
Numerical evidence of 2D Anderson transition for two particles obeying the Anderson-Hubbard model in two dimensions
 was first reported~\cite{Ortugno:AndLocTIPDeloc:EPL99}  on the basis of transmission-amplitude calculations~\cite{McKinnonKramer:TransferMatrix:ZPB83} performed on
 rectangular strips of length  $L=62$ and variable width up to $M=10$. For a pair with zero total energy and for interaction strength $U=1$, the delocalization transition was found to occur for $W=9.3\pm 0.5$.
  The result was also confirmed~\cite{Cuevas:PRL1999} from the analysis of the  energy-level statistics, although with slightly different numbers.  

The existence of a 2D mobility edge for the pair was also reported in Ref.~\cite{Roemer1999}, where a decimation method was employed to compute the critical disorder strength as a function of the interaction strength $U$, based on lattices of similar sizes.
For $U=1.59$, a pair with zero total energy was shown to undergo an Anderson transition at $W=9\pm 0.13$.

Below we verify the existence of the  2D delocalization transition of the pair, following the procedure developed in Ref.~\cite{Stellin:PRB2019}. In order to compare with the previous  numerical  predictions, we set $E=0$ and $W=9$. 
We consider a rectangular strip of dimensions $L, M$, with $L\gg M$, containing  $N=ML$ lattice sites. In order to minimize finite-size effects, the boundary conditions on the single-particle Hamiltonian $H^{sp}$ are chosen periodic in the orthogonal  direction ($y$) and open along the transmission  axis ($x$).
We rewrite the 
rhs of Eq.~(\ref{KE0}) as 
 \begin{equation}\label{KE0bis}
K_{\mathbf n  \mathbf m} =\sum_{r=1}\phi_{\mathbf n r}  \phi_{\mathbf m r}^* \langle \mathbf{n}|G^{\mathrm{sp}}(E-\varepsilon_{r})|\mathbf{m}\rangle,
\end{equation}
 where   $G^{\mathrm{sp}}(\varepsilon)=(\varepsilon I - H^{\mathrm{sp}})^{-1}$  is the Green's function (e.g. the resolvent) of the single-particle Anderson Hamiltonian (\ref{Anderson3D}), $I$ being the identity matrix. 
Due to the open boundary conditions along the longitudinal direction, the Anderson
Hamiltonian possesses a block tridiagonal structure, each block corresponding
to a transverse section of the grid. This structure can be exploited to efficiently compute the 
Green's function $G^{\mathrm{sp}}(\varepsilon)$  in Eq.~(\ref{KE0bis}) via matrix inversion.
In this way the 
total number of elementary operations needed to compute the matrix $K$ scales as $M^{4}L^{3}$, instead of $M^{4}L^{4}$, as naively expected from  the rhs of Eq.~(\ref{KE0}).

Once computed the matrix $K$ of the effective model, we use it to evaluate the logarithm of the transmission amplitude between two transverse sections of the strip as a function of their relative distance $n_x$:
\be\label{logT}
F(n_x)=\ln \sum_{ m_y,n_y} |\langle 1,m_y| G^{\textrm p}(\lambda )|  n_x,n_y \rangle  |^2.
\ee
In Eq.~(\ref{logT}) $G^{\textrm p}(\lambda)=(\lambda I -K)^{-1}$ is the Green's function 
associated to $K$ with $\lambda=1/U$ and the sum is taken over the sites $m_y,n_{y}$ of the two transverse sections.
\begin{figure}
	\centering
	\includegraphics[width=\columnwidth]{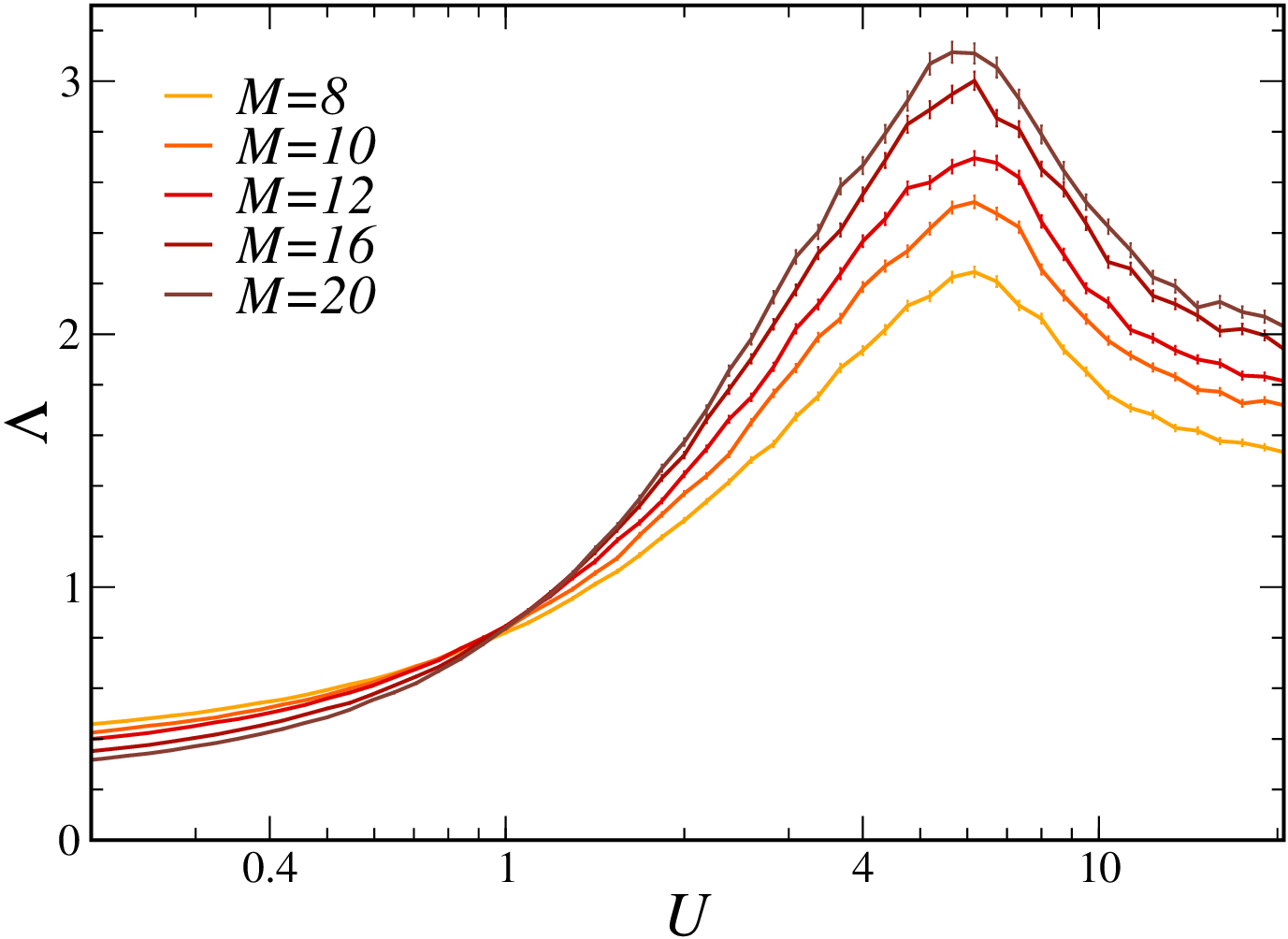}
	\caption{ Reduced localization length of the pair plotted as a function of the interaction strength for increasing  values of the transverse size $M=8, 10, 12, 16, 20$ of the grid. The results are obtained by averaging over $N_{tr}$ different disorder realizations, varying from $N_{tr}=600\; (M=8)$ to $N_{tr}=1000\; (M=20)$.  The disorder strength is fixed to $W=9$ and the pair has zero total energy, $E=0$, 
implying that $\Lambda(-U)=\Lambda(U)$.
	The different curves cross in the interval $0.75<U<1.1$, indicating a possible 2D delocalization transition, as claimed in previous investigations~\cite{Ortugno:AndLocTIPDeloc:EPL99,Roemer1999}. The 2D Anderson transition is actually a finite-size effect, as the crossing points disappear for larger values of $M$, see Fig.\ref{fig:TIP_2D_U-LM_HighM}.}  
	\label{fig:TIP_2D_U-LM_SmallM}
\end{figure}
For each disorder realization, we evaluate $F(n_x)$  at regular intervals along the bar and apply  a linear fit to the data,  $f_{fit}(n_x)=p n_x+q$. For a given value of the interaction strength, we evaluate the (disorder-averaged) Lyapunov exponent $\gamma=\gamma(M,U)$ as $\gamma=-\overline{p}/2$, where $\overline{p}$ is the average of the slope. 
We then infer the localization properties of the system from the behavior of the reduced localization length, which is defined as  $\Lambda=(M \gamma)^{-1}$. In the metallic phase $\Lambda$ increases as $M$ increases, whereas in the insulating phase the opposite trend is seen. At the critical point, $\Lambda$ becomes constant for values of $M$ sufficiently large. Hence the critical point $U=U_c$ of the Anderson transition can be identified by plotting the reduced localization length versus $U$ for different values of the transverse size $M$ and looking at their common crossing points.

In Fig. \ref{fig:TIP_2D_U-LM_SmallM} we show the reduced localization length 
$\Lambda$ as a function of the interaction strength for increasing values of
the strip width, ranging from $M=8$ to $M=20$. The length
of the grid is fixed to $L=400$. Notice that, since $E=0$, the reduced localization length is an even function of the interaction strength, 
$\Lambda(-U)=\Lambda(U)$.
We see  that  $\Lambda$ exhibits a nonmonotonic dependence on $U$, as previously found 
in  one~\cite{Frahm:EigStructAL1DTIP16} and in three~\cite{Stellin:PRB2019} dimensions. In particular, interactions favor the 
delocalization of the pair, the effect being more pronounced near $U=6$.  
We also notice from Fig. \ref{fig:TIP_2D_U-LM_SmallM} that the curves corresponding to  different values of $M$ intersect each others around $U=1$, suggesting a possible phase transition, as previously reported  in Ref.~\cite{Ortugno:AndLocTIPDeloc:EPL99,Roemer1999}. A closer inspection of the data, however, reveals that the crossing points are spread out in the interval  $0.73 \lesssim U \lesssim 1.1$; in particular, they drift to stronger interactions as the system size increases, in analogy with the three-dimensional case~\cite{Stellin:PRB2019}.

\begin{figure}
	\centering
	\includegraphics[width=\columnwidth]{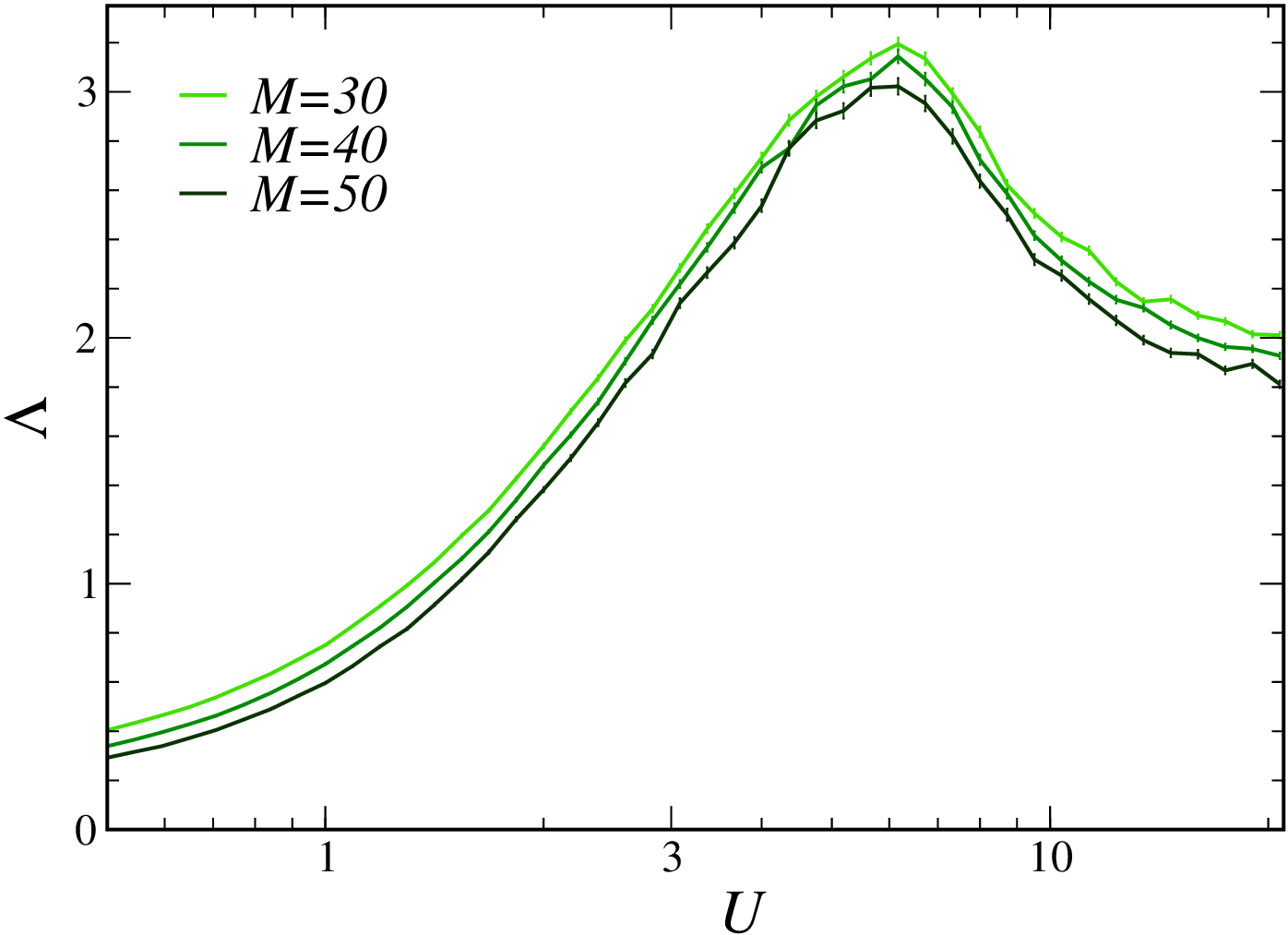}
	\caption{Same plot as in Fig.\ref{fig:TIP_2D_U-LM_SmallM} but for larger grids with  transverse sizes $M=30, 40, 50$ 
	obtained by averaging over $N_{tr}=3600\; (M=30), 4400\; (M=40)$, and $N_{tr}=2850\; (M=50)$ different disorder realizations.	
Notice that all crossing points have disappeared, indicating that the pair is ultimately localized by the disorder for any value of the
interaction strength.}
	\label{fig:TIP_2D_U-LM_HighM}
\end{figure}

A key question is whether a further increase of the strip's width $M$  will only cause a (possibly large) shift of the critical point, or 
rather, the localized phase will ultimately take over for any value of the interaction strength. To  answer this question, we have performed additional calculations using larger grids, corresponding to $M=30, 40, 50$. In order to guarantee a sufficiently large aspect ratio, the 
 length of the bar was fixed to $L=500$. The obtained results are displayed in Fig.\ref{fig:TIP_2D_U-LM_HighM}.
We notice that the crossing points have completely disappeared and the pair localizes in an infinite lattice irrespectively of the specific value of $U$. 
This leads us to conclude that the results of Refs.~\cite{Ortugno:AndLocTIPDeloc:EPL99,Roemer1999}  were plagued by severe finite-size effects, due to the limited computational 
ressources, and no Anderson transition can actually take place for a pair in a disordered lattice of infinite size. 


\section{Pair localization length}
\label{sec:loclength}

Although the pair cannot fully delocalize in two dimensions,
interactions can lead to a drastic enhancement of the two-particle localization length. 
This quantity can be estimated using the one-parameter scaling 
ansatz $\Lambda=f(\tilde \xi/M)$, stating that the reduced localization length
depends solely on the ratio between two quantities: the width $M$ of the strip and a characteristic length $\tilde \xi=\tilde \xi(U,W,E)$, which instead depends on the model parameters and on the total energy of the pair (but not on the system sizes $L, M$). This latter quantity coincides, up to a multiplicative numerical constant $a$, with the  pair localization length, $\xi=a\tilde \xi$. 

We test the scaling ansatz for our effective model (\ref{integral}) using the numerical data for $M=30,40, 50$ displayed in Fig.\ref{fig:TIP_2D_U-LM_HighM}, corresponding to the largest system sizes. 
Let $U_j$, with $j=1,2 ..N_U$, be the values of the interaction strength
used to compute the reduced localization length (in our case $N_U=44$). 
We then determine   the corresponding unknown parameters $\tilde \xi(U=U_j)$ through a least squares procedure, 
following the procedure developed in Ref.~\cite{McKinnonKramer:TransferMatrix:ZPB83}. 
Plotting our data in the form $\ln \Lambda(M,U)$ vs $\ln M$ results in multiple data curves, each of them containing three data points connected by straight lines (corresponding to linear interpolation).
Let $\Lambda_i$ be one of the $(3N_U)$ numerical values available for the reduced localization length. The horizontal line $\ln \Lambda=\ln \Lambda_i$ will generally intersect some of these curves. We find convenient to introduce 
a matrix $\eta$ which keeps track of such events: if the curve $U=U_j$ is crossed,
we set $\eta_{ij}=1$ and call $\ln M_{ij}$ the corresponding point; otherwise  we set $\eta_{ij}=0$.
The unknown parameters are then obtained by minimizing the variance of the difference $\ln M-\ln \tilde \xi$, yielding the
following set of equations (see Ref.~\cite{McKinnonKramer:TransferMatrix:ZPB83} for a detailed derivation): 
\begin{multline}
\label{eqn:scaling}
\sum_{j}\left [\sum_{i}\eta_{ij}\biggl(\frac{1}{N_{i}^{2}}-\frac{\delta_{jk}}{N_{i}}\biggr)\right ]\ln{\tilde \xi (U_{j})}=\\=\sum_{j}\left [\sum_{i}\eta_{ij}\biggl(\frac{1}{N_{i}^{2}}-\frac{\delta_{jk}}{N_{i}}\biggr) \ln M_{ij} \hspace{0.1cm}  \right ] ,
\end{multline}
where $N_i=\sum_j \eta_{ij}$ is the total number of crossing points obtained for each $\Lambda_i$ value.
Equation (\ref{eqn:scaling}) is of the form $AX=B$ and can be easily solved. Notice however that the solution is not unique because
 the matrix $A$ is singular. Indeed the correlation length $\tilde \xi(U)$ is defined up to a multiplicative  constant,
$\tilde \xi\rightarrow a \tilde \xi$, implying that $\ln \tilde \xi$ is defined up to an \emph{additive} 
constant, $\ln \tilde \xi \rightarrow \ln \tilde \xi +\ln a$.

\begin{figure}
	\includegraphics[width=\columnwidth]{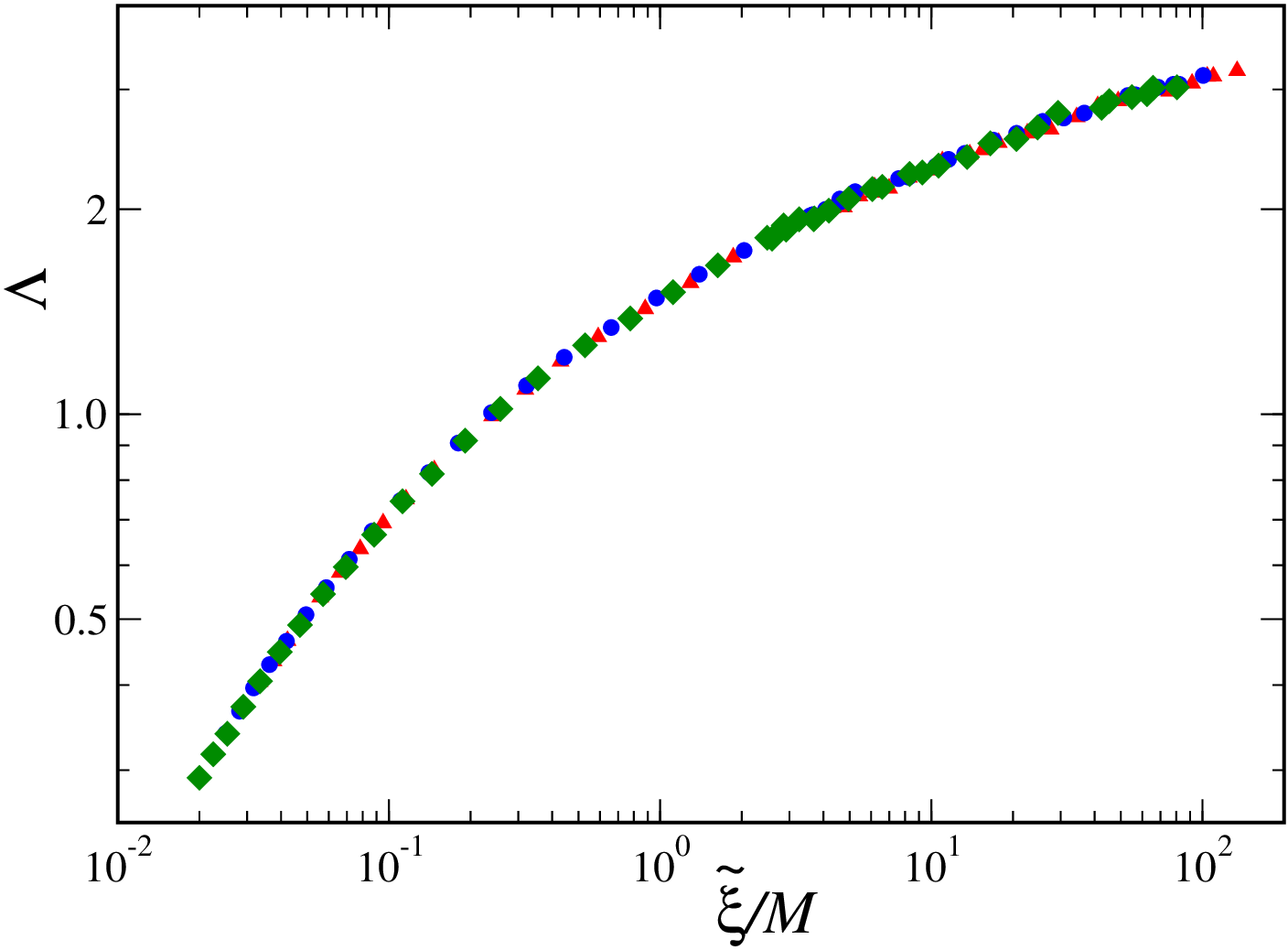}
	\caption{ Double logarithmic plot of the reduced localization length 
		as a function of the ratio $\tilde \xi/M$, where $\tilde \xi$ is the unnormalized localization length obtained from the solution of Eq.~(\ref{eqn:scaling}) and $M$ is the width of the strip.
		 The different symbols correspond to the data for $M=30$ (up triangles), $M=40$ (circles) and $M=50$ (diamonds), shown in Fig.~\ref{fig:TIP_2D_U-LM_HighM}. All data approximately collapse on a single curve, verifying  the scaling ansatz $\Lambda=f(\tilde \xi/M)$.  }
	\label{fig:TIP_2D_xiInvM-LM}
\end{figure}

\begin{figure}
	\includegraphics[width=\columnwidth]{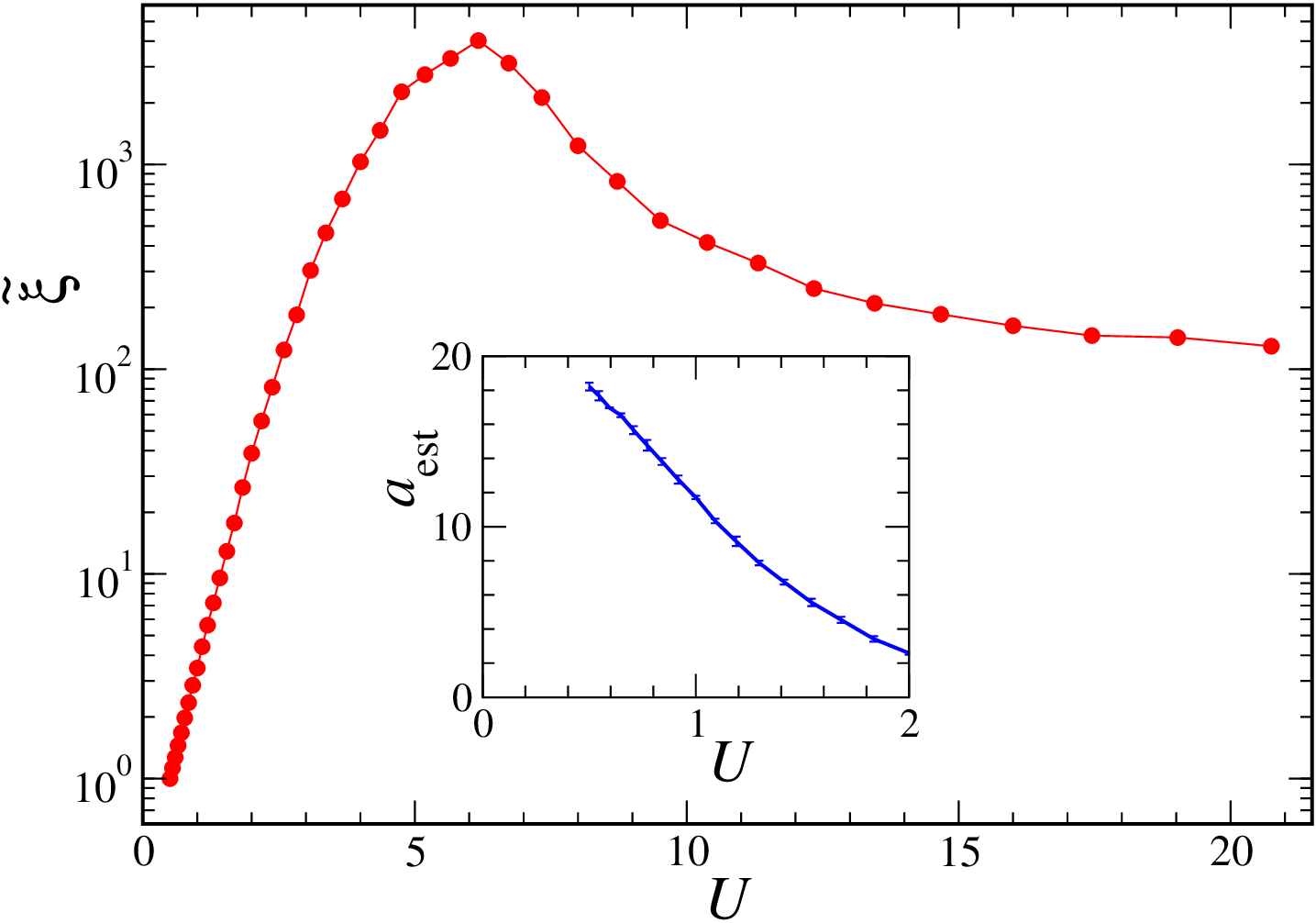}
	\caption{Unnormalized localization length $\tilde \xi$ of the pair plotted as a function of the interaction strength. 
	    Notice the logarithmic scale in the $y$ axis, showing 
		that interactions can enhance the 2D localization length of the pair by more than three orders of magnitude.  The inset displays the estimate  of the multiplicative constant $a$, fixing the absolute scale of the localization length, plotted as a function of the interaction strength. The estimate is obtained by fitting the numerical data in Fig.\ref{fig:TIP_2D_U-LM_HighM} corresponding to weak interactions  using Eq.~(\ref{eqn:finda}), from which we extract $a_\textrm{est}=\xi/\tilde \xi$.  This quantity keeps increasing as $U$ diminishes, signaling that the strongly localized regime is not fully reached in our simulations. 
	 }
	\label{fig:TIP_2D_xi-U}
\end{figure}

In Fig.\ref{fig:TIP_2D_xiInvM-LM} we verify the correctness of the scaling ansatz, by plotting the reduced localization length as a function of the ratio 
$\tilde \xi/M$, where $\tilde \xi$ is obtained from the solution of Eq.~(\ref{eqn:scaling}). We see that our numerical data for different values of the interaction strength and system size do collapse on a single curve, thus confirming the scaling hypothesis.
In the main panel of Fig. \ref{fig:TIP_2D_xi-U} we plot the unnormalized localization length of the pair as a function of the interaction strength. We see that $\tilde \xi$ varies over more than three orders of magnitude in the interval of $U$ values considered. 
In particular, for weak interactions the growth is approximately exponential in $U$, as highlighted by the semi-logarithmic plot. 
Based on analytical arguments, Imry suggested~\cite{Imry:CohPropTIP:EPL95} that the localization length of the pair in the weakly interacting regime should obey the relation $\xi \propto \xi_{\mathrm{sp}}\mathrm{e}^{b(U\xi_{\mathrm{sp}})^{2}}$,
where $\xi_{\mathrm{sp}}$ is the single-particle localization length of the Anderson model and $b$ is a numerical factor. 
A possible reason of the discrepancy is that the cited formula might apply only for relatively modest
 values of the interaction strength, which were not explored in our numerics. 
 Further work will be needed to address this point explicitly. 
 

The constant $a$, allowing to fix the absolute scale of the localization length of the pair,
 is independent of the interaction strength. Its numerical value can in principle be inferred by fitting the data in the strongly localized regime, 
according to
\begin{equation}
\label{eqn:finda}
\Lambda =\frac{\xi}{M}+c\biggl(\frac{\xi}{M}\biggr)^{2},
\end{equation}
where $c$ is a number.  
 In our case the most localized states are those at weak interactions, where the reduced localization length takes its minimum value.  
For each values $U=U_j$ falling in this region, we fit our numerical data according to Eq.~(\ref{eqn:finda}), yielding $\xi=\xi(U)$.
The estimate of the multiplicative constant, which is defined as $a_\textrm{est}=\xi(U)/\tilde \xi (U)$, is displayed  
in the inset of Fig.~\ref{fig:TIP_2D_xi-U}.
Since the estimate of $a$  does not saturates for small $U$, we conclude that, even for the weakest interactions and the largest system sizes considered, the pair has not yet entered the strongly localized regime underlying Eq.~(\ref{eqn:finda}). This asymptotic regime is typically achieved for $\Lambda \lesssim 0.1$, whereas our smallest  value of the reduced localization length is $\Lambda(M=50,U=0.5)\simeq 0.2929$. 
From the inset of Fig.~\ref{fig:TIP_2D_xi-U}  we also see that $a_\textrm{est}$ increases as $U$ diminishes, suggesting that the result obtained for $U=0.5$  actually provides a lower bound for the multiplicative constant. This allows us to conclude that  $a \geq 18.2$. 

\section{Generality of the obtained results}
\label{general}

\begin{figure}
	\includegraphics[width=\columnwidth]{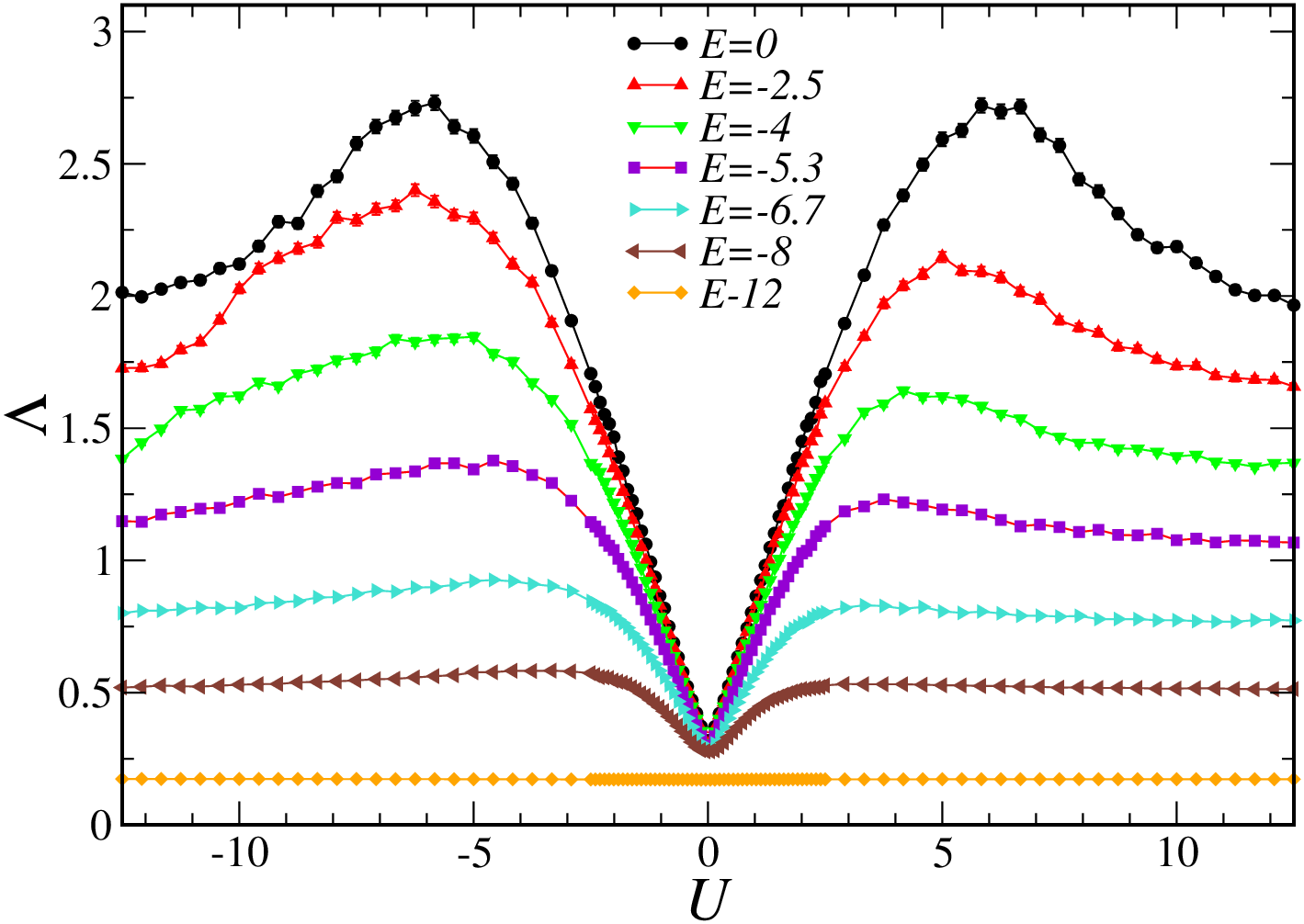}
	\caption{Reduced localization length of the pair as a function of the interaction strength for $W=9$ and for different values of the total energy going from 
	$E=0$ (top curve) to $E=-12$ (bottom curve). The sizes of the strip
	is $M=12$ and $L=400$, while the number of different  disorder realizations is $N_{tr}=1000$.
	The data show that the pair state with zero total energy possesses the largest reduced localization length, see Eq.~(\ref{nonzeroE}), implying that for $W=9$ the pair remains localized for any nonzero total energy.  }
	\label{fig:Efinite}
\end{figure}

In Sec.~\ref{sec:absence} we have  shown that all pair states with total energy $E=0$ are localized for $W=9$. A natural question is whether 
the localization scenario changes at nonzero energy or at weak disorder. 
Let us consider the two cases separately. 
Our  numerical results indicate that, for any values of  $U,W$ and system size $M$, the reduced localization length always takes its maximum value for $E=0$:\begin{equation}\label{nonzeroE}
\Lambda (E,M,U,W)\leq \Lambda(0,M,U,W).
\end{equation}
As an example, in Fig.\ref{fig:Efinite} we plot  $\Lambda$ as a function of the interaction strength, 
for $W=9$ and for different negative values of the energy  (results for positive energies are simply obtained from 
the corresponding data at energy $-E$ by reversing the sign of the interaction strength, $U\rightarrow -U$). 
All calculations are performed on a strip with constant sizes $M=12$ and $L=400$. 
When combined with the finite-size scaling analysis,  the  inequality~(\ref{nonzeroE}) implies that the pair remains localized  for \emph{any} nonzero energy 
with an even shorter localization length, thus excluding a delocalization transition. 
The above inequality expresses the general fact that  the pair can better spread when its total energy lies in the middle of the noninteracting two-particle 
energy spectrum.  For instance, in three dimensions, where genuine Anderson transitions for the pair do occur, we found~\cite{stellin2020twobody}  that 
 metallic regions in the 
interaction-disorder plane become progressively insulating as the energy of the pair departs from zero.  

We note from Fig.\ref{fig:Efinite} that all data curves with $|E|\leq 8$ have absolute minimum at $U=0$. Moreover, the largest enhancement of the reduced localization length 
takes place for weaker interactions as $|E|$ increases. These are specific features of scattering states, whose energy lies inside the noninteracting two-body energy spectrum,
as already observed in one~\cite{Frahm:EigStructAL1DTIP16} and in three~\cite{stellin2020twobody} dimensions. 
 In the asymptotic regime $|E|\gg W$, pairs behave as pointlike molecules and the effective model $K$ takes the form of a single-particle Anderson model, 
as discussed in Ref.~\cite{stellin2020twobody},  which again precludes the possibility of a delocalization transition in two dimensions.

 
Let us now discuss whether an Anderson transition for the pair can appear for weak disorder at fixed total energy, $E=0$. The effective single-particle model $K$ possesses both time reversal and spin rotational symmetries, suggesting that $K$ belongs to the same (orthogonal) universality class of the Anderson model $\hat H^\textrm{sp}$. 
In Ref.~\cite{Stellin:PRB2019} we showed numerically that, in three dimensions, the Anderson transition for a pair with zero energy yields critical exponents in agreement with the predictions of the orthogonal  class. 
Since 2D Anderson transitions are generally forbidden in the orthogonal class, one expects that the pair is localized for \emph{any} finite disorder. For this reason, 
the previous claims of  2D delocalization transitions for two particles are puzzling. Our numerics shows explicitly that these results were 
biased by strong finite-size effects and there is no evidence of violation of the conventional localization scenario.  

\begin{figure}
	\includegraphics[width=\columnwidth]{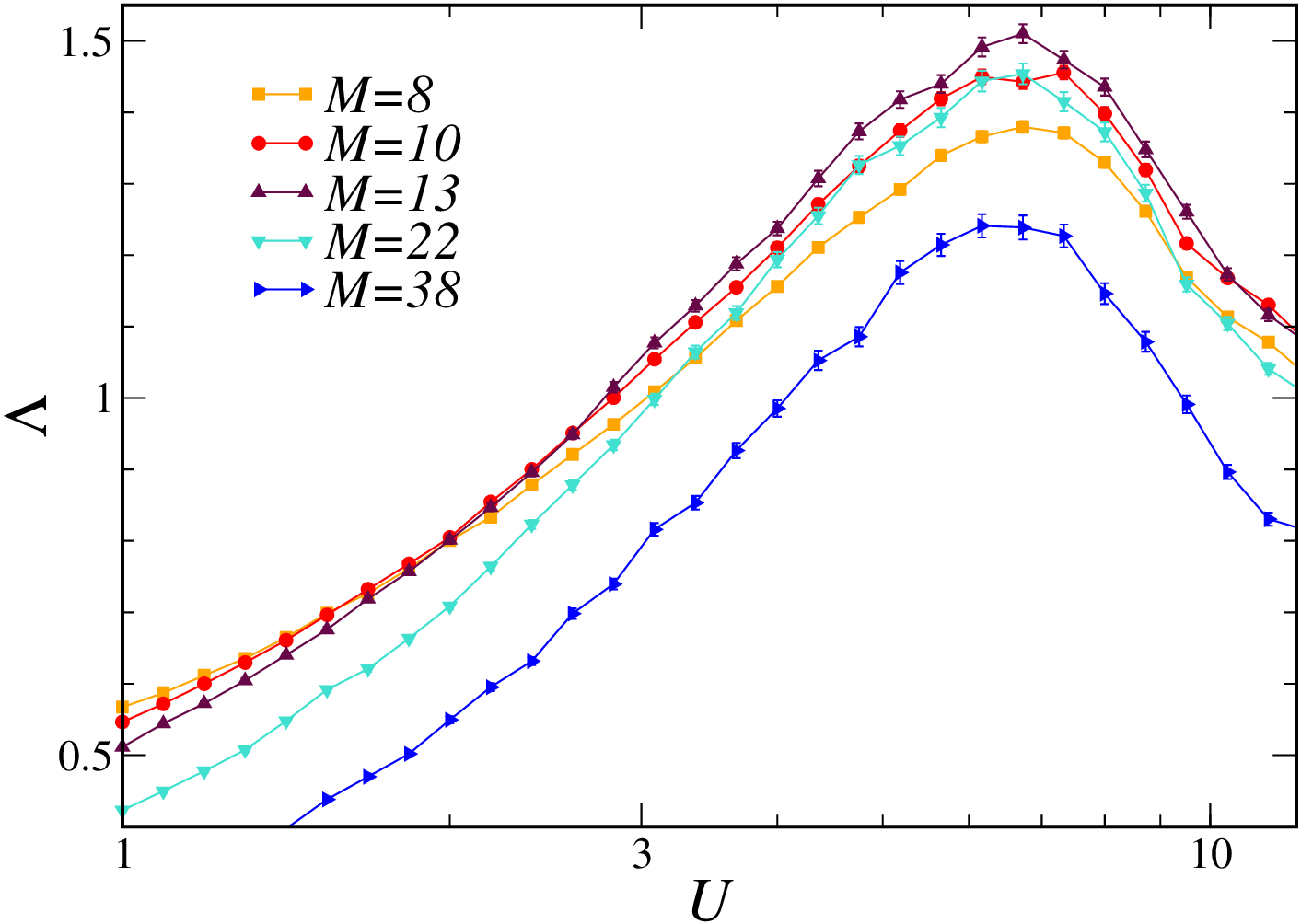}
	\caption{Finite-size scaling analysis for $W=10$ and $E=0$. The reduced localization length is plotted as a function of the interaction strength
	for different system sizes $M=8$ (squares), $10$ (circles), $13$ (up triangles), $22$ (down triangles), and $38$ (right triangles). The length of the strip is $L=400$ for $M\leq 13$
	and $L=500$ otherwise. Notice that the two-particle system exhibits an insulating
	 behavior already for $M=22$. The number of different disorder realizations is $N_{tr}=600$ for $M=38$ and $N_{tr}=1000$ otherwise. }
	\label{fig:W=10}
\end{figure}

From the numerical point of view, the observation of the asymptotic 2D scaling behavior for $W=9$ required large system sizes as compared to
the 3D case studied in Ref.~\cite{Stellin:PRB2019}, where the finite-size scaling analysis was limited to system sizes up to $M=17$. 
Verifying numerically the absence of the 2D transition for weaker disorder is  very challenging,  because
the reduced localization length will exhibit an apparent crossing for even larger  values of $M$ as $W$ diminishes. To appreciate this point, we have repeated the 
same finite-size scaling analysis for $W=10$ and plotted the results in Fig.\ref{fig:W=10}. We see that, already for $M=22$, the pair is localized for any values of the interaction strength, whereas for $W=9$ 
the same asymptotic behavior is reached for larger system sizes, between $M=30$ and $M=40$.

\section{Conclusion and outlook}
\label{sec:conclusions}
Based on an efficient mapping of the two-body Schrodinger equation, we have addressed the localization properties of two bosons or two spin 1/2 fermions in a singlet state obeying the 2D Anderson-Hubbard model.
We have found that no interaction-induced Anderson transition occurs for disordered lattices of infinite size in contrast with previous numerical works, which we have shown to be biased by finite-size effects. In this way we 
reconcile the numerics with the one-parameter scaling theory of localization, predicting the absence of one-particle Anderson transition in two dimensions, in the presence of both time reversal and spin rotational symmetries. Moreover, we found that the pair localization length exhibits a nonmonotonic behavior as a function of $U$, characterized by an exponential 
growth for weak  interactions.  

We point out that the absence of the 2D mobility edge for the two-particle system has been proven for the case of contact interactions; similar conclusions should apply also for short but finite-range interactions. The case of true long-range (e.g Coulomb) interactions is conceptually different and can lead to opposite conclusions~\cite{Cuevas:PRL1999,Shepelyanski:PRB2000}.
From the above discussion, we also expect that  the 2D delocalization transition will appear when the two particles are exposed to spin-orbit couplings, driving the system towards the symplectic universality class,  where  single-particle metal-insulator transitions are generally allowed  even in two dimensions~\cite{Evers:AndersonTransitions:RMP08}.

An interesting and compelling problem is to investigate the implications of our results for a 2D system at finite density of particles, where many-body delocalization transitions have instead been observed, both numerically and experimentally, in  the strongly interacting regime.
We expect that, in the zero density limit, the many-body mobility edge disappears,  irrespective of the bosonic or fermionic 
statistics of the two particles. 
Another interesting direction is to generalize our numerical approach to study the effect of disorder on the transport and spectral properties of excitons in 2D 
semiconductors~\cite{C9CP04111G}.

\section*{ACKNOWLEDGEMENTS} 
We acknowledge D. Delande, K. Frahm, C. Monthus, S. Skipetrov and T. Roscilde for fruitful discussions.
This project has received funding from the European Union's Horizon 2020 research and innovation programme under the 
Marie Sklodowska-Curie grant agreement No 665850. This work was granted access to the HPC resources of CINES (Centre Informatique National de l'Enseignement Sup\' erieur) under the allocations 2017-A0020507629, 2018-A0040507629, 2019-A0060507629 and 2020-A0080507629 supplied by GENCI (Grand Equipement National de Calcul Intensif).

\bibliographystyle{apsrev}
\bibliography{biblio2bodynew2D.bib}

\end{document}